\documentclass[prd,preprintnumbers,twocolumn,amsmath,nofootinbib,amssymb]{revtex4}
\usepackage{graphicx,color,dcolumn,booktabs,bm}
\usepackage{longtable,lscape}
\usepackage{txfonts}
\usepackage{overpic}
\usepackage{amssymb}
\usepackage{epstopdf}
\usepackage{indentfirst}
\usepackage{feynmf}   
\usepackage{slashed}  
\usepackage{cases}
\usepackage{color}
\usepackage{float}
\usepackage{multirow}
\usepackage{ulem}
\usepackage{graphicx,color,dcolumn,booktabs,bm}
\usepackage{epsfig,dsfont,amssymb,amsmath,amsfonts,amsbsy,mathrsfs}

\graphicspath{{Figures/}} %

\makeatletter
\@addtoreset{equation}{section}
\makeatother

\allowdisplaybreaks

\begin{document}

\title{Analysis of the $e^+ e^- \to J/\psi  D\bar D$ reaction close to the threshold concerning claims of a $\chi_{c0}(2P)$ state}

\author{En Wang}
\email{wangen@zzu.edu.cn}
\affiliation{Department of Physics, Guangxi Normal University, Guilin 541004, China}
\affiliation{School of Physics and Microelectronics, Zhengzhou University, Zhengzhou, Henan 450001, China}

\author{Wei-Hong Liang}
\email{liangwh@gxnu.edu.cn}
\affiliation{Department of Physics, Guangxi Normal University, Guilin 541004, China}
\affiliation{Guangxi Key Laboratory of Nuclear Physics and Technology, Guangxi Normal University, Guilin 541004, China}

\author{Eulogio Oset}
\email{eulogio.oset@ific.uv.es}
\affiliation{Department of Physics, Guangxi Normal University, Guilin 541004, China}
\affiliation{Departamento de F\'{i}sica Te\'{o}rica and IFIC, Centro Mixto Universidad de Valencia - CSIC,
Institutos de Investigaci\'{o}n de Paterna, Aptdo. 22085, 46071 Valencia, Spain}

\begin{abstract}
We analyze the $D\bar{D}$ mass distribution from a recent Belle experiment on the $e^+e^- \to J/\psi D\bar{D}$ reaction, and show that the mass distribution divided by phase space does not have a clear peak above the $D\bar{D}$ threshold that justifies the experimental claim of a $\chi_{c0}(2P)$ state from those data.
Then we use a unitary formalism with coupled channels $D^+D^-$, $D^0\bar{D}^0$, $D_s\bar{D}_s$, and $\eta\eta$, with some of the interactions taken from a theoretical model, and use the data to fix other parameters.
We then show that, given the poor quality of the data, we can get different fits leading to very different $D\bar{D}$ amplitudes, some of them supporting a $D\bar{D}$ bound state and others not. The main conclusion is that the claim for the $\chi_{c0}(2P)$ state, already included in the PDG, is premature, but refined data can provide very valuable information on the $D\bar{D}$ scattering amplitude. As side effects, we warn about the use of  a Breit-Wigner amplitude parameterization close to threshold, and show that the $D_s\bar{D}_s$ channel  plays an important role in this reaction.
\end{abstract}

\maketitle



\section{Introduction}
\label{sec:intro}

A recent experiment by the Belle Collaboration on the $ e^+ e^- \to J/\psi  D\bar D$ reaction, looking into the $D\bar D$ mass distribution, observes a broad peak close to threshold~\cite{exp}.
A fit to the data in terms of a Breit-Wigner amplitude produces a mass and width $M=3862 ^{+26+40}_{-32-13} \; {\rm MeV}$, $\Gamma=201 ^{+154+88}_{-67-82} \; {\rm MeV}$.
The $J^{PC}=0^{++}$ hypothesis is favored over the $2^{++}$.
The peak is associated to a new charmonium state $X(3860)$ which the authors propose as a candidate for $\chi_{c0}(2P)$.
In the present work we will show that these data are not precise enough to support the existence of a resonance state around 3860~MeV.

The issue of the $\chi_{c0}(2P)$ state has a long story. The charmonium $\chi_{c0}(1P)$ state, with $I^G (J^{PC})= 0^+ (0^{++})$, is the well known $\chi_{c0}(1P)$ at 3415 MeV tabulated in the PDG~\cite{Tanabashi:2018oca}. The state $\chi_{c0}(2P)$ has been predicted in the Godfrey-Isgur relativized potential quark model~\cite{Barnes:2005pb} to be around 3916 MeV. This fact, together with the discovery of the $X(3915)$ state at Belle~\cite{Uehara:2009tx,Shen:2009vs} in $\gamma \gamma \to J/\psi \omega$, prompted early suggestions to associate the  $X(3915)$ to the missing $\chi_{c0}(2P)$ state~\cite{Liu:2009fe}. This, together with the study done at BaBar~\cite{Lees:2012xs}, suggesting that the spin and parity of the $X(3915)$ were $0^+$,  had as a consequence the official introduction of the $X(3915)$ state as the $\chi_{c0}(2P)$ in the PDG of 2013, where it stayed till 2016 as a consequence of a study~\cite{Zhou:2015uva} which suggested that the $X(3915)$ state could be the same as $X(3930)$ with quantum numbers $2^{++}$. This latter scenario is also supported in Ref.~\cite{Yu:2017bsj} and Ref.~\cite{Ortega:2017qmg}, where it is suggested that the $X(3915)$ and $X(3930)$ resonances arise as different decay mechanisms of the same $2^{++}$ state. The hypothesis of Ref.~\cite{Liu:2009fe} was also challenged in Refs.~\cite{Guo:2012tv,Olsen:2014maa} on the basis of the large width that the $\chi_{c0}(2P)$
should have decaying into the $D \bar{D}$ channel (see for instance Refs.~\cite{Barnes:2006xq,Yang:2009fj}), which has not been observed for the $X(3915)$, and possible inconsistencies in the mass compared to other charmonium states. Actually, some quark models get a reduced width to $D\bar{D}$ of the radially excited $\chi_{c0}(2P)$ state because of the node in the wave function~\cite{Jiang:2013epa}, and the enhancement of the OZI suppressed width into $J/\psi \omega$ can also find an explanation in a generalized screened potential model~\cite{Gonzalez:2016fsr}. The $\chi_{c0}(2P)$ hypothesis for the $X(3915)$ has been reinforced by new studies in Refs.~\cite{Chen:2013yxa,Duan:2020tsx}. In Ref.~\cite{Chen:2013yxa} assuming that the $X(3915)$ and $X(3930)$ are the $\chi_{c0}(2P)$ and $\chi_{c2}(2P)$ respectively, it is shown that the decay
width of $X(3930) \to J/\psi\omega$ is at least one order of magnitude smaller than that of $X(3915)\to J/\psi\omega$, which would justify why only the $X(3915)$ has been observed in the $J/\psi\omega$ invariant mass spectrum for the process $\gamma\gamma \to J/\psi\omega$. In Ref.~\cite{Duan:2020tsx} it is shown that the use of the unquenched quark model, where, in addition to the quark components, the meson-meson components are taken into account via loops with these intermediate states~\cite{Kalashnikova:2005ui,Pennington:2007xr,Zhou:2013ada,Li:2009ad,Ono:1983rd}, provides an answer to the problems posed to the $\chi_{c0}(2P)$ hypothesis in Refs.~\cite{Guo:2012tv,Olsen:2014maa}, in particular the puzzle of the charmonium masses and the decay widths discussed above.
  The former discussion is proper in a paper that discusses the experimental claims for the $\chi_{c0}(2P)$ state in  Ref.~\cite{exp}, and a  further discussion of these issues is given in Ref.~~\cite{Duan:2020tsx}, but it serves the purpose to show that the question of the assignment of the $\chi_{c0}(2P)$ state to some of the observed states or experimental peaks is the object of an intense debate at present.

The reason to reopen the issue is that the analysis of a structure close to threshold, as the one seen in the experiment of Ref.~\cite{exp}, demands techniques that are consistent with analyticity, unitarity in coupled channels and threshold properties (cusps in amplitudes) that go beyond the simple representation by a Breit-Wigner amplitude~\cite{Hanhart,Guorep}.
As an example, in Ref.~\cite{alberto} a peak observed close to threshold of the $\phi \omega$ mass distribution in the $J/\psi \to \gamma \phi \omega$ reaction \cite{exp2},
associated there to a state with mass around 1795 MeV and width around 95 MeV (with large uncertainties),
was found to be compatible with the effect of the $f_0(1710)$,
which is below the $\phi \omega$ threshold.

In order to investigate the potential information of the present data, we use a dynamical model with coupled channels and unitarity, including the $D^+D^-$, $D^0\bar{D}^0$, $D_s\bar{D}_s$ channels, plus the $\eta\eta$ channel as a means to account for the contribution of light channels to this process. We fit some parameters of the model to the data, resulting in quite different $D\bar{D}$ amplitudes, indicating that the data are not precise enough to come with a unique answer.

In particular, dividing the cross section data by phase space we clearly observe  that there is no peak structure that justifies the claim for the 3862~MeV state in Ref.~\cite{exp}. The detailed study also serves us to show the danger of parameterizing amplitudes by a Breit-Wigner form  close to threshold, and the role that the $D_s\bar{D}_s$ channel has  in the interpretation of a clear cusp at this threshold.

The study done here lays the grounds to obtain precise information on the $D\bar{D}$ scattering amplitudes from improved data and this should serve to encourage the experimentalists to give that step in the future.

\section{Formalism}
\label{sec:form}

\begin{figure}[b!]
\begin{center}
\includegraphics[scale=0.6]{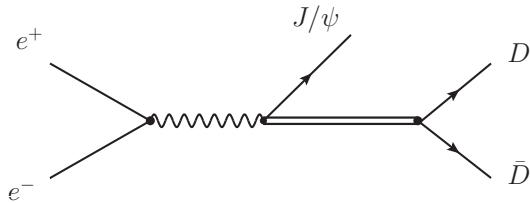}
\end{center}
\caption{Diagrammatic representation of the $e^+ e^- \to J/\psi D\bar D$ reaction.}
\label{Fig:1}
\end{figure}
Ignoring factors which depend on the energy $\sqrt{s}$ of the $e^+ e^-$ system or are constant, we can write the $D\bar D$ mass distribution of the reaction as \cite{daniexp}
\begin{equation}\label{eq:dsigma}
  \frac{{\rm d} \sigma}{{\rm d} M_{\rm inv}(D\bar D)} = \mathcal{C} \;\frac{1}{(2\pi)^3}\; \frac{m_e^2}{s\sqrt{s}}\; |\vec{p}\,|\; |\tilde{k}|\; |T|^2,
\end{equation}
where $\vec p$ is the $J/\psi$ momentum in the $e^+ e^-$ center of mass frame and $\tilde k$ the $D$ momentum in the $D\bar D$ rest frame,
\begin{eqnarray}
|\vec p\,| & = & \frac{\lambda^{1/2}(s, M^2_{J/\psi}, M^2_{\rm inv}(D\bar D))}{2 \sqrt{s}}, \\
|\tilde k|   & = & \frac{\lambda^{1/2}(M^2_{\rm inv}(D\bar D), M^2_D, M^2_{\bar D})}{2\,M_{\rm inv}(D\bar D)}.
\end{eqnarray}

\begin{figure}[tbhp]
\begin{center}
\includegraphics[scale=0.6]{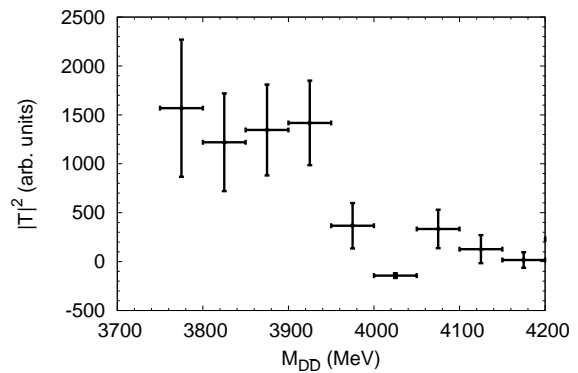}
\end{center}
\caption{Data for $\frac{{\rm d} \sigma}{{\rm d} M_{\rm inv}(D\bar D)}$ divided by the phase space of Eq.~\eqref{eq:dsigma}~\cite{exp}.}
\label{Fig:2}
\end{figure}

The $ e^+ e^- \to J/\psi  D\bar D$ process is depicted in Fig. \ref{Fig:1}.
The experiment of Ref. \cite{exp} is done around the $\Upsilon(1S)$ to $\Upsilon(5S)$ states, hence $\sqrt{s}$ for $e^+ e^-$ ranges from 9.46 GeV to 10.87 GeV.
The value of $|\vec p\, |$ is smoothly dependent on $M_{\rm inv}(D\bar D)$ in this range and we take $\sqrt{s}=10 \;{\rm GeV}$ for the calculations.
The magnitude $T$ appearing in Eq. \eqref{eq:dsigma} is the $D\bar D \to D\bar D$ amplitude to which we come below,
but before elaborating on it, we find most instructive to show the results for $|T|^2$ obtained from the data, dividing the experimental cross section by the phase space factor of Eq.~\eqref{eq:dsigma}, $|\vec{p}\,|\; |\tilde{k}|$.
The results are shown in Fig.~\ref{Fig:2}, where the experimental data are taken from Fig.~6 of Ref.~\cite{exp}, from where the data of the background shown in Fig.~5 of Ref.~\cite{exp} has been subtracted.~\footnote{We thank K. Chikilin for informing us that this is the appropriate procedure. We should also note that the data presented in this published paper are not corrected by acceptance, hence, some caution must be taken on the conclusions.}

What we see in Fig.~\ref{Fig:2} is that in the region around 3860~MeV where the $\chi_{c0}(2P)$ was claimed (yet, with a large width of about 200~MeV), there is no peak that justifies the existence of a state. We should note that there is an extra experimental point close to threshold, but the bin of $50$~MeV does not allow one to get a meaningful value for the phase space.  We should also call the attention to the fact that the sharp fall down of the data in Fig.~\ref{Fig:2}, corresponds exactly to the $D_s \bar{D}_s$ threshold, where a cusp should in principle be expected.
It should also be noted that this region is around 3937~MeV where another resonance, the $\chi_{c2}(3930)$, appears. This resonance is narrow ($\Gamma \approx 35$~MeV) and couples to $D\bar{D}$ as clearly seen in the $\gamma\gamma \to D\bar{D}$ experiments of Belle~\cite{Uehara:2005qd} and BaBar~\cite{Aubert:2010ab}, and in $pp$ collisions at LHCb~\cite{Aaij:2019evc}. One can appreciate that the quality of the data is rather poor and the peak, clearly seen in other experiments, is not visible here, nor was claimed in the experimental analysis of Ref.~\cite{exp}.
 We shall come back to this point below.

After this observation, let us present our analysis. Following Ref. \cite{Gamermann}, we construct the $D\bar D$ amplitude using the Bethe-Salpeter equation in coupled channels,
\begin{equation}\label{eq:BSE}
  T=[1-VG]^{-1}\, V,
\end{equation}
with the channels $D^+D^-$, $D^0 \bar D^0$, $D_s \bar D_s$, and $\eta \eta$, where $V_{ij}$ are the transition potentials and $G_i$ the diagonal matrix accounting for the two-meson loop function for each channel.
The important channels for the threshold behaviour are $D^+ D^-$ and $D^0 \bar D^0$.
In Ref. \cite{Gamermann}, in addition to the $D\bar D, D_s \bar D_s$ channels, other light pseudoscalar-pseudoscalar ($PP$) channels are considered, including $\pi \pi, K\bar K, \eta \eta$.
Their couplings to the $D\bar D$ channels are very much suppressed and their roles around the $D\bar D$ threshold are negligible~\footnote{One should note that in Ref.~\cite{Gamermann} the $\eta$ was considered as an SU(3) octet, as in most works using chiral Lagrangians. However, the peculiar nature of the $D\bar{D}$ bound state as an SU(3) singlet made the coupling to the $\eta\eta$ channel very small. In Ref.~\cite{Gamermann:2008jh} the standard $\eta-\eta'$ mixing of octet and singlet~\cite{Bramon:1992kr} was assumed, as a consequence of which the couplings to $\eta\eta$, $\eta\eta'$, and $\eta'\eta'$ became sizeable, although still small compared to those to $D\bar{D}$ and $D_s\bar{D}_s$.}.

The only effect of these light $PP$ channels is to provide a small imaginary part to the $D\bar{D}$ amplitude below threshold, and, thus, a width to a potential $D\bar{D}$ bound state.
Due to this, in Ref.~\cite{Dai} all light channels considered in Ref.~\cite{Gamermann}, that led to a width of the $D\bar D$ state of about 36 MeV,
were integrated in just one channel, the $\eta \eta$, and the transition potential from $D\bar D \to \eta \eta$ was tuned such as to give that width.
Here we follow the same strategy but take this transition potential as a free parameter.
Then, as in Ref.~\cite{Dai} we take the $V_{ij}$ matrix elements between $D$ and $D_s$ from Ref. \cite{Gamermann} and $V_{D^+ D^-, \eta \eta}= V_{D^0 \bar D^0, \eta \eta}=a$, $V_{D_s \bar D_s, \eta \eta}= V_{\eta \eta, \eta \eta}=0$.

It should be stressed that, Eq.~(\ref{eq:BSE}), factorizing the $D \bar{D}\to D \bar{D}$ $T$ matrix, relies upon the Migdal-Watson approximation, which is justified close to threshold when the scattering amplitude is large~\cite{Watson:1952ji,migdal,Hanhart:2003pg,Hanhart:1998rn}, as it is expected here where there is a bound state close to threshold~\cite{Gamermann}. It is often used in the literature~\cite{Sibirtsev:2004id,daniexp} and improvements when this condition is not fulfilled are also considered in the literature~\cite{Baru:2000hg,Haidenbauer:2005eh,Hanhart:1998rn,Oset:2016lyh}.

For the $G$ function of the rest of the channels, we use dimensional regularization as in Ref.~\cite{Gamermann}, but with the scale mass $\mu$ fixed to $\mu=1500$ MeV, and the subtraction constant $\alpha$, common to the $D^+ D^-, D^0 \bar D^0, D_s \bar D_s$ channels, as a free parameter.
For $\alpha=-1.3$, a $D\bar{D}$ bound state was found in Ref.~\cite{Gamermann}.
For the $\eta\eta$ channel, we use the same subtraction constant as in Ref.~\cite{Gamermann}. The real part of $G_{\eta\eta}$ plays a negligible role in the results. We have checked that using only the imaginary part,
\begin{equation}\label{eq:ImG}
  i\, {\rm Im}\, G_{\eta \eta} (M_{\rm inv}) = - \frac{i}{8\pi}\; \frac{1}{M_{\rm inv}}\; q_{\eta},
\end{equation}
with $q_{\eta}= \lambda^{1/2}(M^2_{\rm inv}, m^2_\eta, m^2_\eta)/2 M_{\rm inv}$, gives results undistinguishable to the sight, which we will show in the next section.

\section{Results}
\label{sec:res}

\begin{figure}[tbhp]
\begin{center}
\includegraphics[scale=0.65]{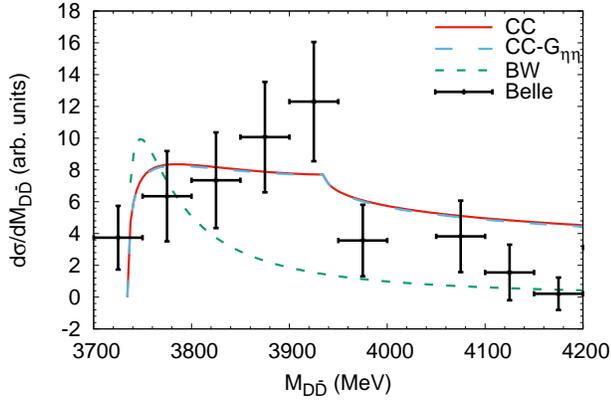}
\end{center}
\caption{The differential cross section of the reaction $e^+e^-\to J/\psi D \bar{D}$. The line labeled as `CC' is the result obtained in the coupled channel
method, where only the imaginary part of $G_{\eta\eta}$ is taken, and the line labeled as `CC-$G_{\eta\eta}$' corresponds to one where both the real and imaginary parts of $G_{\eta\eta}$ are taken.
The curve labeled as `BW' shows the results of a Breit-Wigner form, where $M_X = 3706$~MeV, and $\Gamma_X = 50$~MeV. The parameters are $a=50$ and $\alpha=-1.3$.}
\label{Fig:dcs_3par}
\end{figure}

\begin{figure}[tbhp]
\begin{center}
\includegraphics[scale=0.55]{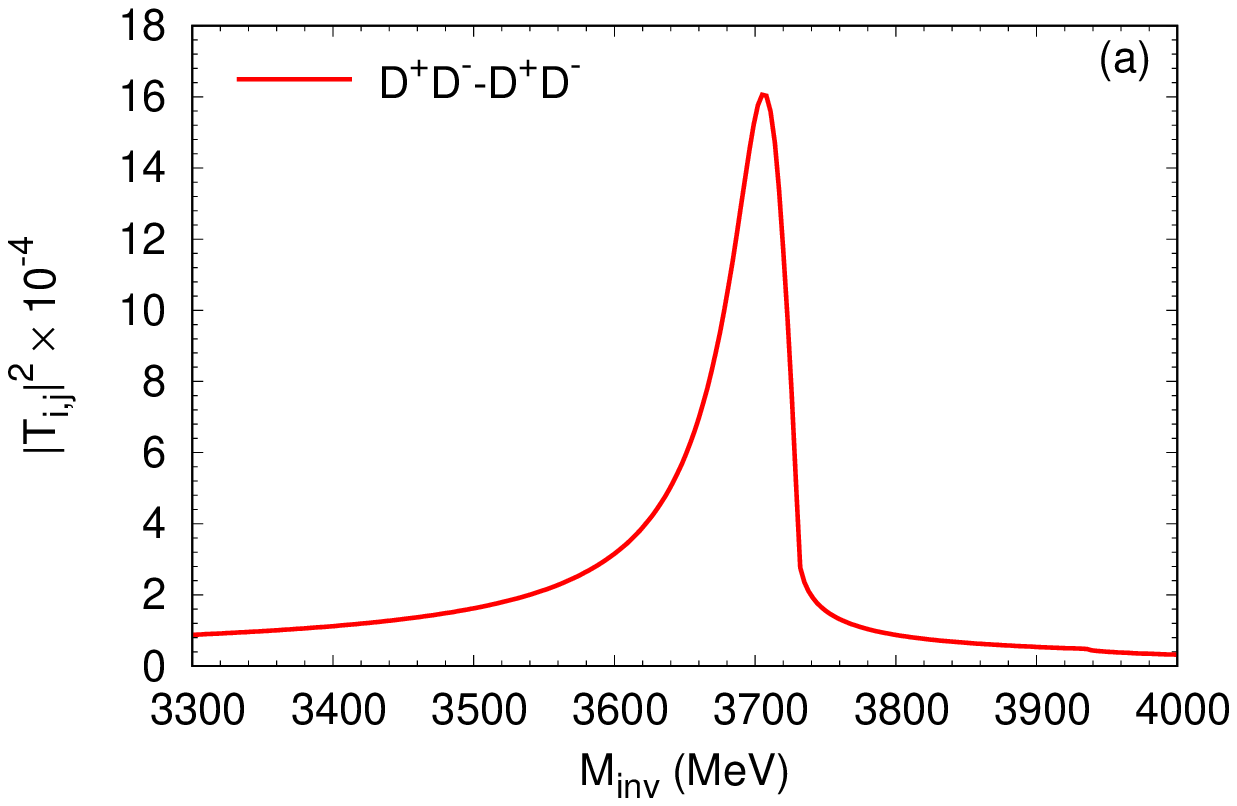}
\includegraphics[scale=0.55]{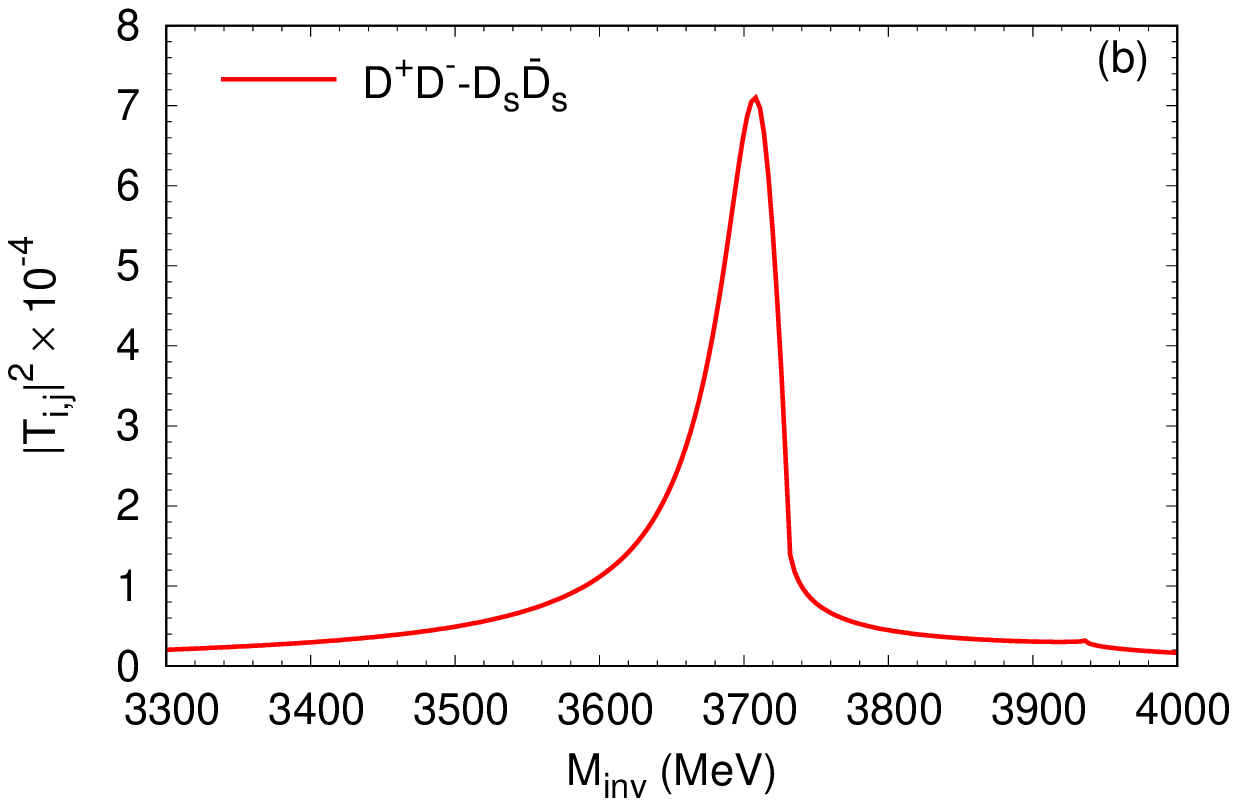}
\includegraphics[scale=0.55]{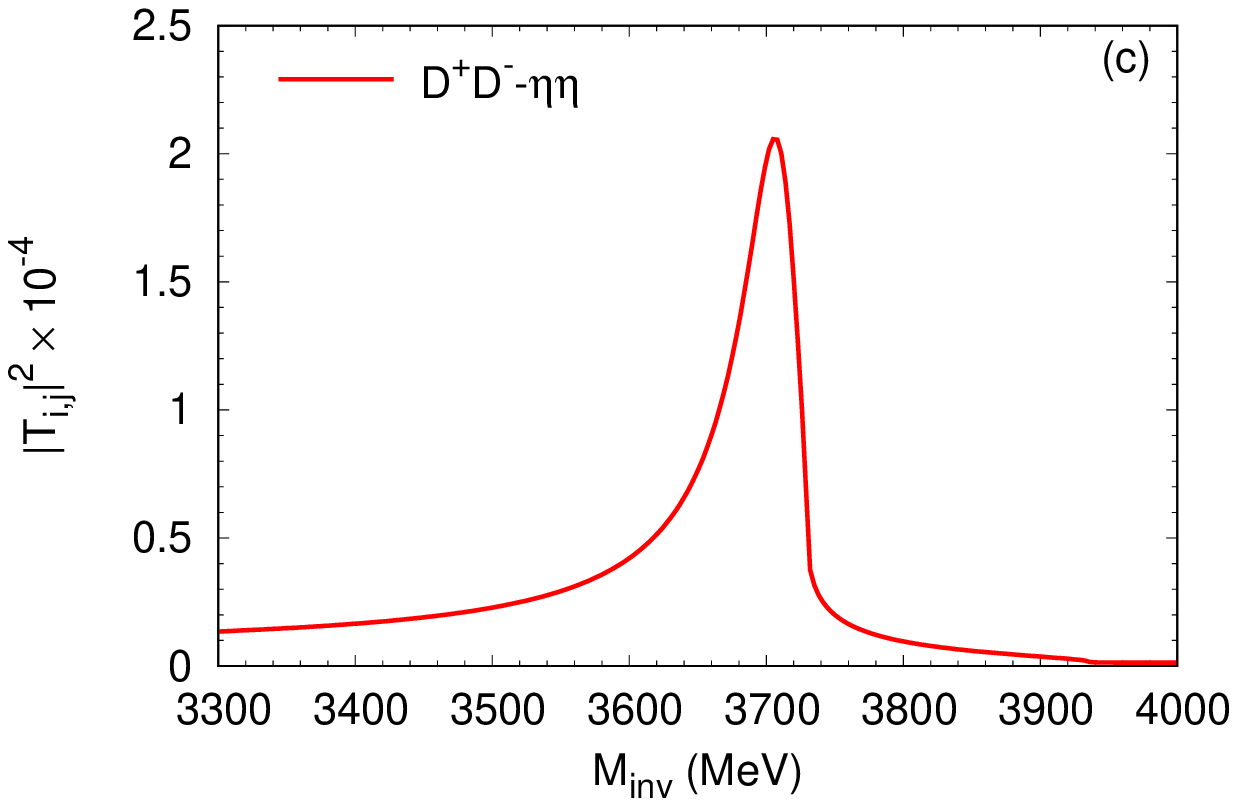}
\end{center}
\caption{(Color online) The modulus squared of the amplitudes: (a), $|T_{D^+D^-\to D^+D^-}|^2$, (b), $|T_{D^+D^-\to D_s\bar{D}_s}|^2$, and (c),  $|T_{D^+D^-\to \eta\eta}|^2$. The parameters are $a=50$ and $\alpha=-1.3$.} 
\label{Fig:amp_3par}
\end{figure}

The procedure followed has three free parameters, the constant $\mathcal{C}$ in Eq. \eqref{eq:dsigma}, the transition potential $a$ between $D\bar{D}$ and $\eta\eta$, and the subtraction constant $\alpha$. The amplitudes that our model produces have a limited range of validity and should not be used much above the $D_s\bar{D}_s$ threshold.
It should be noted that we rely only on $D\bar{D}$ $S$-wave interaction and hence we cannot obtain the $\chi_{c2}(3930)$ resonance which couples to $D\bar{D}$ in $D$-wave.
 There are few experimental points in that range, with large errors and furthermore there is the handicap of not having the acceptance  corrected data.
It is then not surprising that we can get many different fits to the data.
Given the few parameters of the theory and the limited range of our formalism, we consider for the fit the points below $M_{\rm inv}=3900$~MeV. This leaves four data points, and the lowest energy one is not very indicative given the large binning of the experiment. One is then left with three significative points to make the fit, which gives us much freedom and this is why we choose values of $\alpha$ in a range of values around those used in other analyses. The idea is to inspect how much freedom  we have to obtain a reasonable agreement with the data. An interesting side comment on this approach is that, even fitting only the lowest invariant mass data, the results obtained for large invariant masses are not too bad compared with experiment, particularly when compared with just phase space as we shall see below.
 We show three of them that correspond to quite different $D\bar{D}$ amplitudes. In the first place, we fix $\alpha=-1.3$ as in Ref.~\cite{Gamermann} and fit $\mathcal{C}$ and $a$. The parameter $\mathcal{C}$ fixes the global strength and is irrelevant comparing  to data in arbitrary units. The parameter $a$ turns out to be $a=50$. The results for the differential mass distributions are shown in Fig.~\ref{Fig:dcs_3par}.
From the figure, we also find that the real part of the $G_{\eta\eta}$ plays a negligible role, by  comparing the `CC-$G_{\eta\eta}$' curve, where the real part is taken, to the `CC' curve, where the real part is not taken.
 In Fig.~\ref{Fig:amp_3par}, we show the results for $|T|^2$ for different channels, with the $T$ matrix found in Eq.~\eqref{eq:BSE}, and we see that the amplitude corresponds to a $D\bar{D}$ bound state with mass $M_X=3706$~MeV, and width $\Gamma_X=50$~MeV.

\begin{figure}[tbhp]
\begin{center}
\includegraphics[scale=0.65]{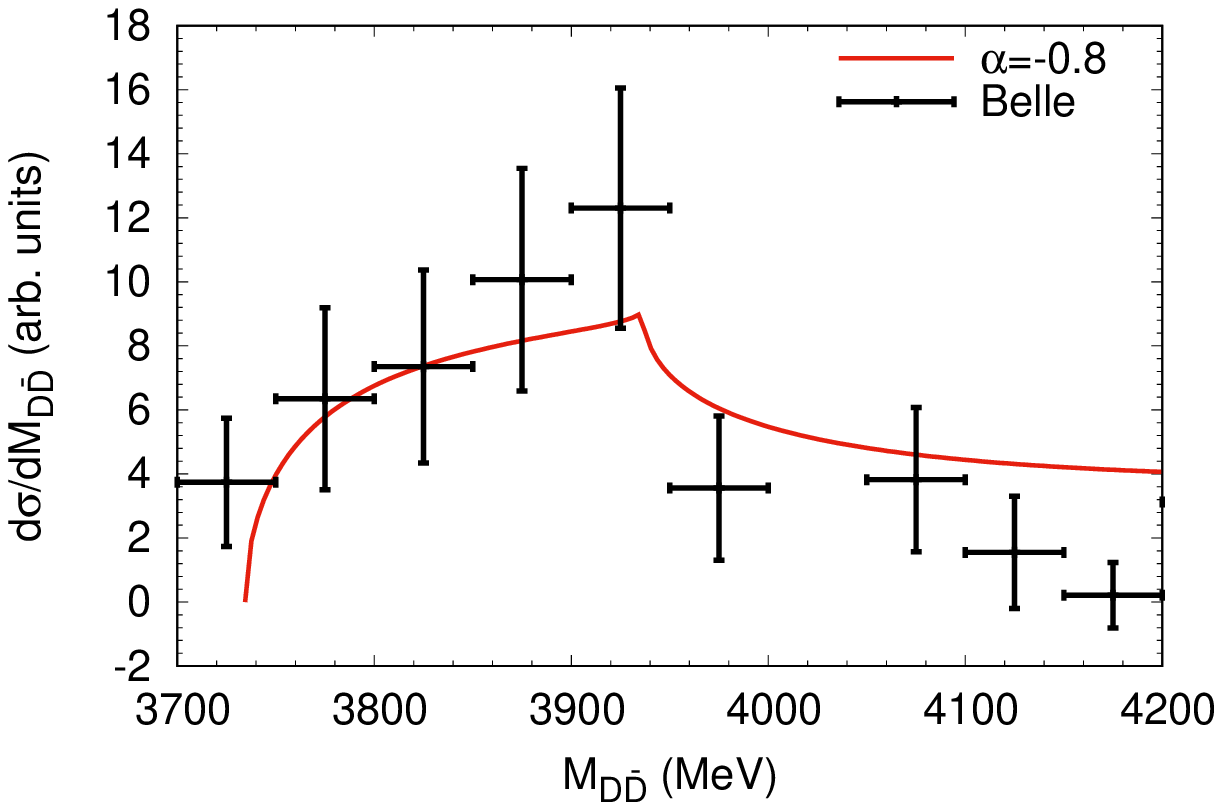}
\end{center}
\caption{$D\bar{D}$ mass distribution for the case of $\alpha=-0.8$ and including the factor 1.6 for the potentials  $V_{D^+D^-,D_s\bar{D}_s}$ and $V_{D^0\bar{D}^0,D_s\bar{D}_s}$. }\label{Fig:result-0.8}
\end{figure}

\begin{figure}[tbhp]
\begin{center}
\includegraphics[scale=0.65]{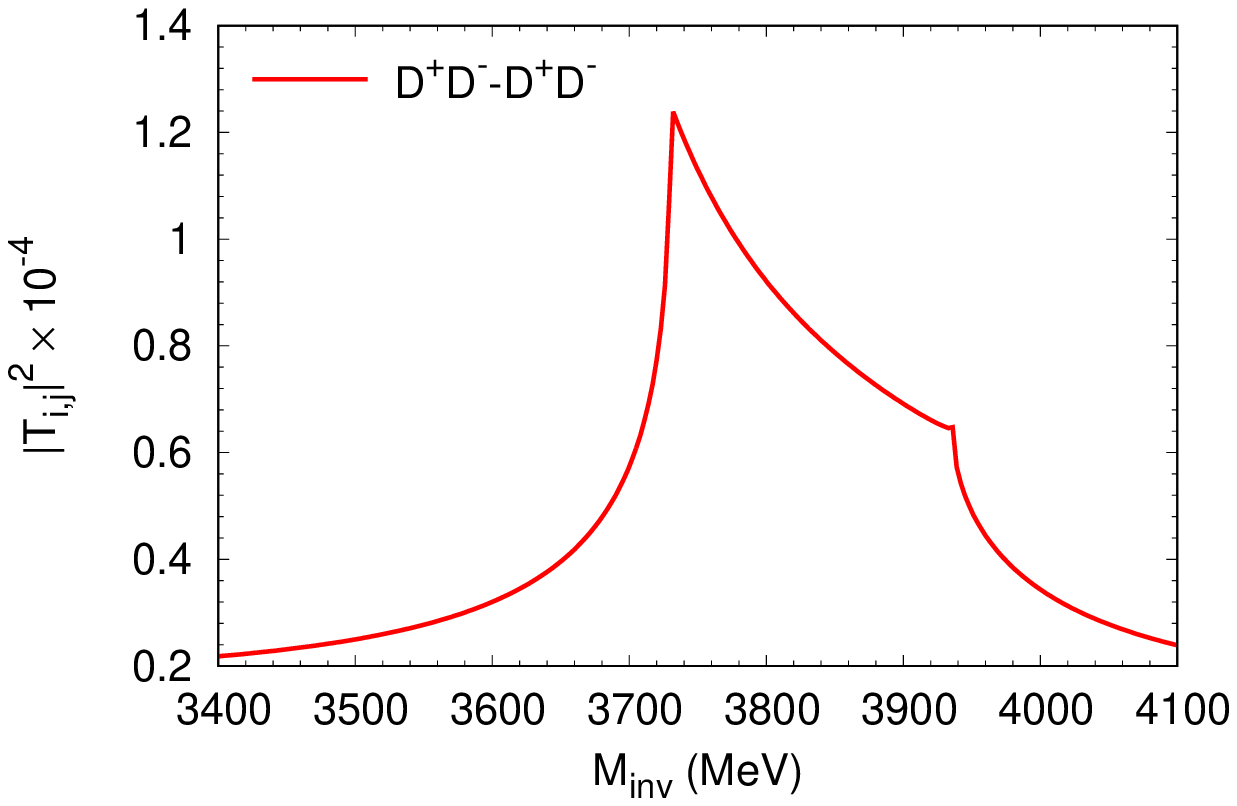}
\end{center}
\caption{Modulus squared of the $D^+D^-\to D^+D^-$ amplitude for the case of $\alpha=-0.8$ and including the factor 1.6 for the potentials  $V_{D^+D^-,D_s\bar{D}_s}$ and $V_{D^0\bar{D}^0,D_s\bar{D}_s}$.}\label{Fig:amp-0.8}
\end{figure}

\begin{figure}[tbhp]
\begin{center}
\includegraphics[scale=0.65]{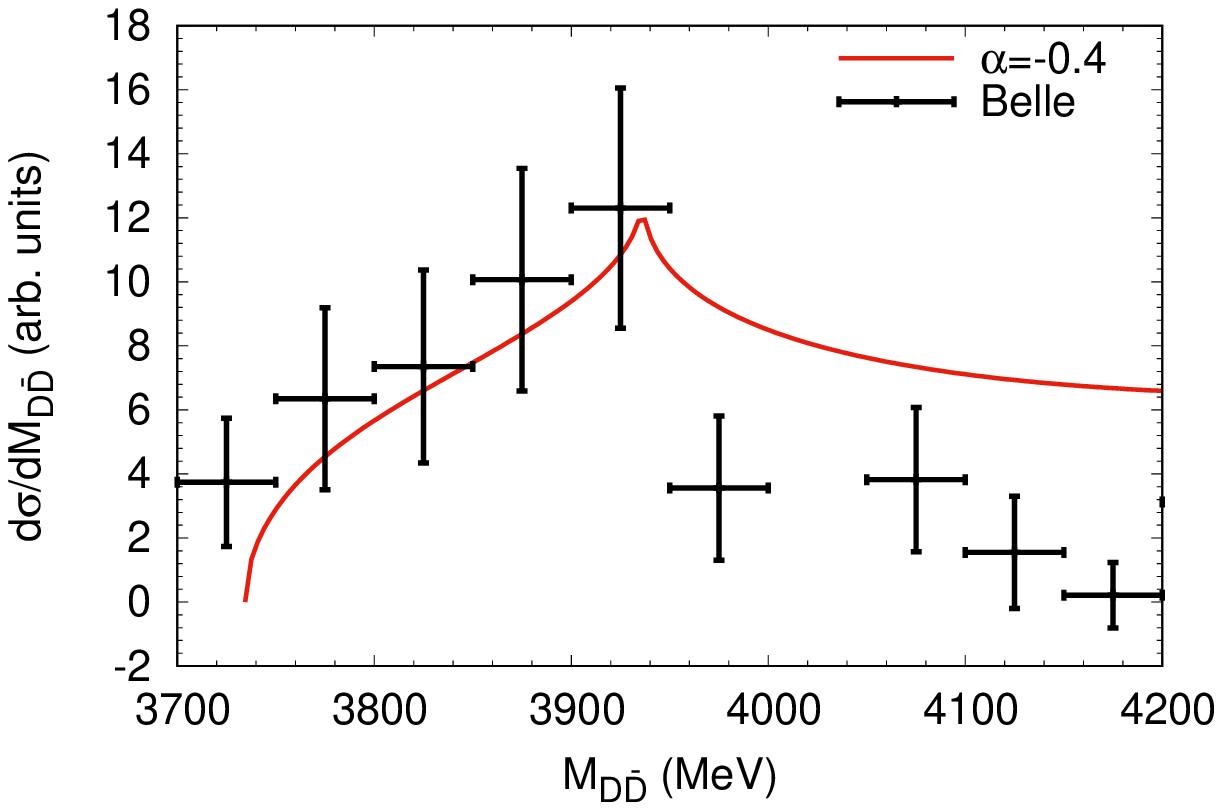}
\end{center}
\caption{$D\bar{D}$ mass distribution for the case of $\alpha=-0.4$ and including the factor 1.45 for the potentials  $V_{D^+D^-,D_s\bar{D}_s}$ and $V_{D^0\bar{D}^0,D_s\bar{D}_s}$.}\label{Fig:dcs-0.4}
\end{figure}

\begin{figure}[tbhp]
\begin{center}
\includegraphics[scale=0.65]{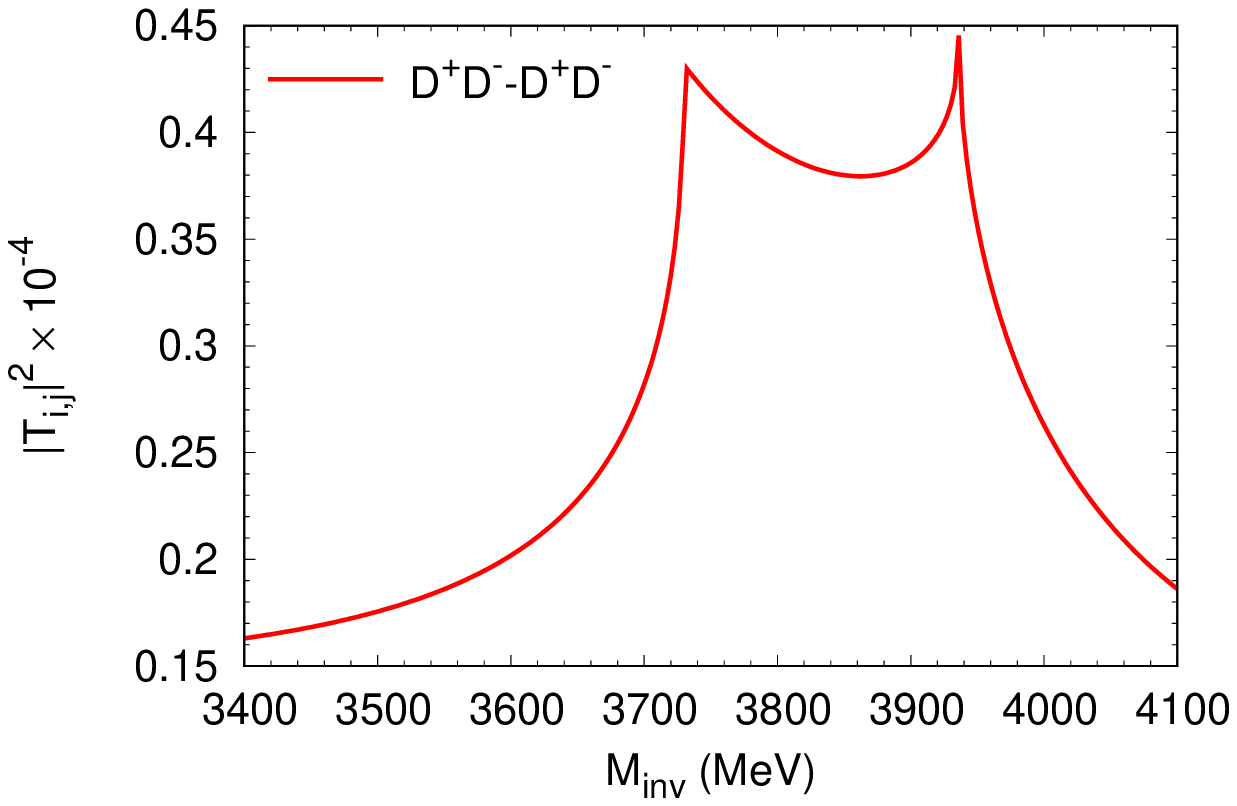}
\end{center}
\caption{Modulus squared of the $D^+D^-\to D^+D^-$ amplitude for the case of $\alpha=-0.4$ and including the factor 1.45 for the potentials  $V_{D^+D^-,D_s\bar{D}_s}$ and $V_{D^0\bar{D}^0,D_s\bar{D}_s}$.}\label{Fig:amp-0.4}
\end{figure}

Next, we take a Breit-Wigner amplitude,
\begin{equation}
T=\frac{\beta}{M^2_{\rm inv}(D\bar{D}) - M^2_{X}+ i M_{X}\Gamma_{X}},
\end{equation}
the parameter $\beta$ gives the strength, and we take $M_{X}=3706$~MeV and $\Gamma_{X}=50$~MeV, the mass and width of the $D\bar{D}$ bound state, as determined previously in the coupled channel approach.

We show the results of the Breit-Wigner amplitude in Fig.~\ref{Fig:dcs_3par} compared to those of the coupled channel approach. We see that the Breit-Wigner amplitude and the coupled channel approach give rise to very different shapes in spite of sharing the same mass and width of the state. This exercise  is very illuminating concerning the use of Breit-Wigner amplitudes close to threshold, something strongly discouraged in Refs.~\cite{Hanhart,Guorep} (see also Refs.~\cite{Hanhart:2015zyp,Hyodo:2020czb,Wang:2020pyy,Wang:2019mph}).  We should also note the strong Flatt\'e effect in the coupled channels amplitudes in Fig.~\ref{Fig:amp_3par}, due to the opening of the $D \bar D$ threshold.
Certainly one can improve the Breit-Wigner amplitude using energy dependent widths as done in Ref.~\cite{exp} and Ref.~\cite{Guo:2012tv}, and even incorporate the Flatt\'e effect allowing the coupling to the $D\bar{D}$ channel, but the effect of having at the same time the two coupled channels $D\bar{D}$ and $D_s\bar{D}_s$ is not easy to incorporate unless an explicit unitary approach in coupled channel is considered, as done here (see a very recent discussion on thresholds and coupled channels in Ref.~\cite{Dong:2020hxe}).

The other comment worth making is that the coupled channel approach, that contains the $D_s\bar{D}_s$ channel explicitly, produces a cusp at the $D_s\bar{D}_s$ threshold, and with the crudeness of the data, there seems to be a clear indication of such a cusp in the experiment.
We should stress that only an approach which contains the $D\bar{D}$ and $D_s\bar{D}_s$ channels explicitly, with a transition potential between the two channels, can give rise to this cusp.
The coupling of the $D\bar{D}$ bound state in this case to $D_s\bar{D}_s$ channel should be found as mostly responsible for the strength of the coupled channel amplitude close to the $D_s\bar{D}_s$ threshold compared to the simplified Breit-Wigner amplitude.
As mentioned above, one should also be careful about taking too seriously the apparent cusp in the experimental data given the fact that there is no peak for the $\chi_{c2}(3930)$ that one would expect in that region. The presence of this state, together with an experimental resolution of 50~MeV, bigger than the $\chi_{c2}(3930)$ width of 35~MeV, could have something to do with the experimental jump seen around the $D_s\bar{D}_s$ threshold, such that not all these effects should be attributed to a $D_s\bar{D}_s$ cusp in $S$-wave.

In Fig.~\ref{Fig:result-0.8}, we make another fit with the value for $\alpha=-0.8$. We can get an acceptable fit to the data to the expense of changing a bit the transition potentials $V_{D^+D^-,D_s\bar{D}_s}$ and $V_{D^0\bar{D}^0,D_s\bar{D}_s}$ of Ref.~\cite{Gamermann}, multiplying them by a factor 1.6, which stresses more the cusp effect. The value of $a$ is 0.07. The interesting thing is that now we do not have a $D\bar{D}$ bound state.  The $D^+D^-\to D^+D^-$ amplitude is shown in Fig.~\ref{Fig:amp-0.8}, where clearly there is no peak in the $D\bar{D}$ bound region, but there are two very distinct cusps at the $D\bar{D}$ and $D_s\bar{D}_s$ thresholds.

We perform yet another fit, having the strength of $\alpha$ smaller, taking this time $\alpha=-0.4$. We can see the result for the mass distribution in Fig.~\ref{Fig:dcs-0.4}. We have also performed another fit by increasing the transition potentials $V_{D^+D^-,D_s\bar{D}_s}$ and $V_{D^0\bar{D}^0,D_s\bar{D}_s}$ by a factor 1.45. The value of $a$ is now $a=0.02$.  
Given the crudeness of the data, the fit results are still acceptable. Here we would like to show how much the shape and strength of $|T|^2$ has changed compared to the former case, which we can see comparing Fig.~\ref{Fig:amp-0.8} and Fig.~\ref{Fig:amp-0.4}.

The three cases studied clearly show how far we are from being able to determine the $D\bar{D}\to D\bar{D}$ amplitude from the present data and the need for more precise data to be able to determine this amplitude  unambiguously. Yet, the potential of this reaction to determine whether there is or not a bound $D\bar{D}$ state should stimulate further experimental work on this reaction.

\begin{figure}[tbhp]
\begin{center}
\includegraphics[scale=0.65]{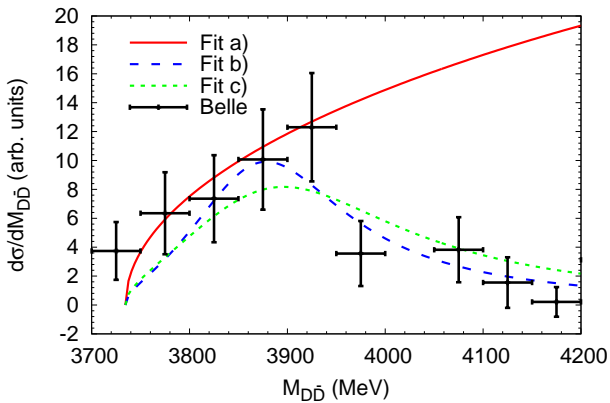}
\end{center}
\caption{The $D\bar{D}$ mass distribution for $e^+e^-\to J/\psi D\bar{D}$. The curves labeled as `Fit a)', `Fit b)', and `Fit c)', correspond to the fit with just phase space, the fit with mass 3862~MeV and the central width 201~MeV, and fit mass 3862~MeV and the width 300~MeV, respectively.}\label{Fig:ds_BW}
\end{figure}

In a different line we perform now another three fits using the conventional approach where only the strength has to  be adjusted: a) fit with just phase space, b) fit with the mass 3862~MeV and the central width 201~MeV of Ref.~\cite{exp}, c) fit with mass 3862~MeV and the width 300~MeV, well within the error bars of $\Gamma=201^{+154+88}_{-67~-82}$~MeV of Ref.~\cite{exp}. The results can be seen in Fig.~\ref{Fig:ds_BW}. 
As one can see, if one looks only to the lowest invariant mass points, the phase space alone gives a fair description. Certainly it deviates grossly from the data as we go above the $D_s \bar{D}_s$ threshold. The fits involving a state with mass 3682 MeV and widths 200~MeV or 300~MeV, produce a qualitative agreement with the data, but a closer look shows that it is not as good as for instance the results of Fig.~\ref{Fig:result-0.8}  at low invariant masses below 3900~MeV, although at higher invariant masses it fits the data better. First, to look for a resonance it is better to look at data divided by phase space, as in Fig.~\ref{Fig:2}, where no obvious peak is devised. Second, the low invariant mass results are not good and the phase space alone looks better. Also, the seeming agreement with the data above the $D_s\bar{D}_s$ threshold should be taken with care after the discussion made above on the role of the $D_s\bar{D}_s$ threshold and the possible contribution in this region of the $\chi_{c2}(3930)$ state, which has not been disentangled in the experimental analysis.  
Actually, in this context it is worth mentioning that in the work of Ref.~\cite{Guo:2012tv} a Breit-Wigner fit was done to the $\gamma\gamma \to D\bar{D}$ data, providing a possible $\chi_{c0}(2P)$ state with a lower mass, 3837~MeV, than Ref.~\cite{exp}. However, the authors were cautious quoting ``{\it More refined analysis of the data with higher statistics is definitely necessary to confirm our assertion}''.
On the other hand one can make the arguments in a different direction, showing that the agreement seen in Fig.~\ref{Fig:result-0.8} with our model at low invariant masses can also be obtained with phase space alone. Only one conclusion is clear from this discussion and it is that the quality of the data is too poor to obtain any conclusion concerning states of any type. Yet, the shapes obtained with the different assumptions are quite different and could be discerned with more data with much smaller errors. The potential of the reaction to provide this information is there but any claims at present with these data are premature.

Finally, it would be interesting to put together several reactions producing $D\bar{D}$ close to threshold and analyzing them together. Some steps in this direction have been given in Ref.~\cite{Wang:2020elp}, where the $\gamma\gamma\to D\bar{D}$ reactions using data from Belle~\cite{Uehara:2005qd} and BaBar~\cite{Aubert:2010ab} have been analyzed, supporting a $D\bar{D}$ bound state in $I=0$, yet with large errors. An extra fit is done in Ref.~\cite{Wang:2020elp}, fitting the $e^+e^-\to J/\psi D\bar{D}$ data studied here together with the $\gamma\gamma\to D\bar{D}$ data. The higher quality of the  $\gamma\gamma\to D\bar{D}$ data makes the effect of the $e^+e^-\to J/\psi D\bar{D}$ data in the fit moderate, but the global fit still prefers a bound $D\bar{D}$ state, yet with errors that do not allow a precise quantification of the position and width. Complementing the message of the present paper, it was also shown in Ref.~\cite{Wang:2020elp} that the data of Belle and BaBar divided by phase space did not show any bump to support a state at 3862 MeV as claimed in Ref.~\cite{exp} and mildly suggested in Ref.~\cite{Guo:2012tv}.

We should note that a $D\bar{D}$ bound state, in analogy to the $K\bar{K}$ bound state that stands for the $f_0(980)$~\cite{isgur,npa,Kaiser,Markushin,Nieves}, was predicted in Refs.~\cite{Gamermann,Nieves:2012tt,HidalgoDuque:2012pq}. 
A boost to this idea has been given recently with the finding of a $D\bar{D}$ bound state in the Lattice QCD calculation of Ref.~\cite{Prelovsek:2020eiw} with a binding of $\mathcal{B}=4.0^{+5.0}_{-3.7}$~MeV. Simultaneously the LHCb Collaboration has started a thorough program studying the $B\to D\bar{D}h$, with $h$ a different hadron, where a large  strength is observed at the $D\bar{D}$ threshold~\cite{ddbarh}. Although an explanation was given in Ref.~\cite{ddbarh} in base to the decay of the $\psi(3770)$ resonance into $D\bar{D}$, the advent of the results of Ref.~\cite{Prelovsek:2020eiw} has motivated the experimental team to measure the threshold region with more precision to investigate a possible signal  of a bound $D\bar{D}$ state~\cite{Tim}.

\section{Conclusions}
\label{sec:conc}

We have done a reanalysis of the $e^+e^-\to J/\psi D\bar{D}$ data~\cite{exp}, by looking at the $D\bar{D}$ mass distribution, from where the existence of a new charmonium state $X(3860)$ was claimed~\cite{exp}. This conclusion was based on a fit to the data with a Breit-Wigner structure. However, we argue that structures close to threshold require a more sophisticated treatment, demanding unitarity in coupled channels,  something which has been further stressed in Ref.~\cite{Dong:2020hxe}.

We have performed this work using the channels $D^+D^-$, $D^0\bar{D}^0$, $D_s\bar{D}_s$, and in addition the $\eta\eta$ channel,  used solely as a means of providing a small imaginary part to the $D\bar{D} \to D\bar{D}$ transition below threshold.
 The Bethe-Salpeter equation in coupled channels is evaluated taking the same transition potentials from early work on meson-meson scattering in the charm sector that describes basic phenomenology~\cite{Gamermann}, and fitting the free parameters to the data.

 We can summarize our findings as follows:

 1) The data of Ref.~\cite{exp} divided by phase space did not show any obvious peak which could justify the claims of a $\chi_{c0}(2P)$ state. 
Certainly, the poor quality of the data does not exclude such a possibility, or even the possibility to have a bound $D\bar{D}$ state and another broad state. Yet, the fact that the data does not show the $\chi_{c2}(3930)$, clearly seen in other experiments, makes the interpretation of the results of any fit to data including the region of 3900 MeV to 4000 MeV more questionable.

 2) We clearly showed that a Breit-Wigner amplitude in the case of poles close to threshold is not appropriate  to represent  the results that one gets with a coupled channel unitary approach for energies close to threshold. A general study of this problem, in line with the claims of this point has been recently done in Ref.~\cite{Dong:2020hxe}.

 3) In the same way that we questioned the claim of a state at 3860~MeV made in Ref.~\cite{exp}, we also showed that, even in the dynamical  picture  of unitary coupled channels, the quality of the data  did not allow us to get any conclusions on whether there is or not a $D\bar{D}$ bound state.

4) The analysis also showed that the data, even with the limited accuracy, seems to show the existence of a clear cusp  at the $D_s\bar{D}_s$ threshold. 
However, it could mask the contribution of the $\chi_{c2}(3930)$ which is not identified in the analysis of Ref.~\cite{exp}. In any case, this fact indicates the convenience of making an analysis using coupled  $D\bar{D}$ and $D_s\bar{D}_s$ channels.

 5) The study shows the potential of this reaction to extract information on the existence or not of a  bound $D\bar{D}$ state with more data and more precision around threshold.
\footnote{We take advantage to ask the authors of Ref.~\cite{exp} to provide the corrected data, and in general to all authors of the experimental community. Data should be provided in a way that allows to be contrasted by a theory if we wish to extract valuable scientific information from them.} This should work as an incentive for the experimentalist to look again  into this reaction to provide this improved statistics.

6) The data used here were also taken in connection with the $\gamma\gamma \to D\bar{D}$ data of Belle~\cite{Uehara:2005qd} and BaBar~\cite{Aubert:2010ab}, and a global fit to the data showed a preference for a bound $D\bar{D}$ state very close to the threshold~\cite{Wang:2020elp}, yet with large errors. Ultimately, the information obtained from all these reactions, measured with much better statistics, should bring a definite answer to the question of the $D\bar{D}$ bound state. Yet, concerning the existence of the state at 3862~MeV, the study done in Ref.~\cite{Wang:2020elp} with the $\gamma\gamma \to D\bar{D}$ reaction agrees with the present one, and also disfavors the existence of that state.

NOTE: After this work was completed, a detailed analysis of the $B^+\to D^+D^- h$ ($h$ an extra hadron) reaction has been conducted, leading to the publication of the papers~\cite{Aaij:2020ypa,Aaij:2020hon}. From the invariant mass and angular distributions, two $\chi_{cJ}$ resonances are reported, sitting at the same mass, the $\chi_{c0}(3930)$ and the $\chi_{c2}(3930)$, with widths around 17~MeV and 34~MeV, respectively. Fits with the $\chi_{c0}(3860)$ of Ref.~\cite{exp} are explicitly conducted and found unfavorable, concluding with the statement ``{\it There is no evidence for the $\chi_{c0}(3860)$ state reported by the Belle Collaboration~\cite{exp}}".

\begin{acknowledgments}
This work is partly supported by the National Natural Science Foundation of China under Grants Nos. 11975083, 11947413. 
It is also supported by the Key Research Projects of Henan Higher Education Institutions under No. 20A140027,  the Project of Youth Backbone Teachers of Colleges and Universities of Henan Province (2020GGJS017), the Academic Improvement Project of Zhengzhou University, and the Fundamental Research Cultivation Fund for Young Teachers of Zhengzhou University (JC202041042).
This work is also partly supported by the Spanish Ministerio de Economia y Competitividad
and European FEDER funds under the contract number FIS2011-28853-C02-01, FIS2011-28853-C02-02, FIS2014-57026-REDT, FIS2014-51948-C2-1-P, and FIS2014-51948-C2-2-P.
\end{acknowledgments}



\begin{thebibliography}{99}

\bibitem{exp}
  K.~Chilikin {\it et al.} [Belle Collaboration],
  Observation of an alternative $\chi_{c0}(2P)$ candidate in $e^+ e^- \rightarrow J/\psi D \bar{D}$,
  Phys.\ Rev.\ D {\bf 95}, 112003 (2017).

\bibitem{Tanabashi:2018oca}
  M.~Tanabashi {\it et al.} [Particle Data Group],
  Review of Particle Physics,
  Phys.\ Rev.\ D {\bf 98}, 030001 (2018).


\bibitem{Barnes:2005pb}
T.~Barnes, S.~Godfrey and E.~Swanson,
Higher charmonia,
Phys. Rev. D \textbf{72}, 054026 (2005).

\bibitem{Uehara:2009tx}
S.~Uehara \textit{et al.} [Belle],
Observation of a charmonium-like enhancement in the $\gamma \gamma \to \omega J/\psi$ process,
Phys. Rev. Lett. \textbf{104}, 092001 (2010).

\bibitem{Shen:2009vs}
C.~Shen \textit{et al.} [Belle],
Evidence for a new resonance and search for the $Y(4140)$ in the $\gamma \gamma \to \phi J/\psi$ process,
Phys. Rev. Lett. \textbf{104}, 112004 (2010).

\bibitem{Liu:2009fe}
X.~Liu, Z.~G.~Luo and Z.~F.~Sun,
$X(3915)$ and $X(4350)$ as new members in $P$-wave charmonium family,
Phys. Rev. Lett. \textbf{104}, 122001 (2010).

\bibitem{Lees:2012xs}
J.~Lees \textit{et al.} [BaBar],
Study of $X(3915) \to J/\psi \omega$ in two-photon collisions,
Phys. Rev. D \textbf{86}, 072002 (2012).

\bibitem{Zhou:2015uva}
Z.~Y.~Zhou, Z.~Xiao and H.~Q.~Zhou,
Could the $X(3915)$ and the $X(3930)$ Be the Same Tensor State?,
Phys. Rev. Lett. \textbf{115},  022001 (2015).

\bibitem{Yu:2017bsj}
G.~L.~Yu, Z.~G.~Wang and Z.~Y.~Li,
The analysis of the charmonium-like states $X^{*}(3860)$, $X(3872)$, $X(3915)$, $X(3930)$ and $X(3940)$ according to its strong decay behaviors,
Chin. Phys. C \textbf{42},  043107 (2018).

\bibitem{Ortega:2017qmg}
P.~G.~Ortega, J.~Segovia, D.~R.~Entem and F.~Fern\'andez,
Charmonium resonances in the 3.9 GeV/$c^2$ energy region and the $X(3915)/X(3930)$ puzzle,
Phys. Lett. B \textbf{778}, 1-5 (2018).

\bibitem{Guo:2012tv}
  F.~K.~Guo and U.~G.~Mei{\ss}ner,
  Where is the $\chi_{c0}(2P)$?,
  Phys.\ Rev.\ D {\bf 86}, 091501 (2012).

\bibitem{Olsen:2014maa}
S.~L.~Olsen,
Is the $X(3915)$ the $\chi_{c0}(2P)$?,
Phys. Rev. D \textbf{91}, 057501 (2015).

\bibitem{Barnes:2006xq}
T.~Barnes,
The $XYZ$s of charmonium at BES,
Int. J. Mod. Phys. A \textbf{21}, 5583-5591 (2006).

\bibitem{Yang:2009fj}
Y.~c.~Yang, Z.~Xia and J.~Ping,
Are the $X(4160)$ and $X(3915)$ charmonium states?,
Phys. Rev. D \textbf{81}, 094003 (2010).

\bibitem{Jiang:2013epa}
Y.~Jiang, G.~L.~Wang, T.~Wang and W.~L.~Ju,
Why $X(3915)$ is so narrow as a $\chi_{c0}(2P)$ state?,
Int. J. Mod. Phys. A \textbf{28}, 1350145 (2013).

\bibitem{Gonzalez:2016fsr}
P.~Gonz\'alez,
A quark model study of strong decays of $X\left( 3915\right) $,
J. Phys. G \textbf{44},  075004 (2017).

\bibitem{Chen:2013yxa}
D.~Y.~Chen, X.~Liu and T.~Matsuki,
Hidden-charm decays of $X(3915)$ and $Z(3930)$ as the $P$-wave charmonia,
PTEP \textbf{2015},  043B05 (2015).

\bibitem{Duan:2020tsx}
M.~X.~Duan, S.~Q.~Luo, X.~Liu and T.~Matsuki,
Possibility of charmoniumlike state $X(3915)$ as $\chi_{c0}(2P)$ state,
Phys. Rev. D \textbf{101},  054029 (2020).


\bibitem{Kalashnikova:2005ui}
Y.~Kalashnikova,
Coupled-channel model for charmonium levels and an option for $X(3872)$,
Phys. Rev. D \textbf{72}, 034010 (2005).

\bibitem{Pennington:2007xr}
M.~Pennington and D.~Wilson,
Decay channels and charmonium mass-shifts,
Phys. Rev. D \textbf{76}, 077502 (2007).

\bibitem{Zhou:2013ada}
Z.~Y.~Zhou and Z.~Xiao,
Comprehending heavy charmonia and their decays by hadron loop effects,
Eur. Phys. J. A \textbf{50}, 165 (2014).

\bibitem{Li:2009ad}
B.~Q.~Li, C.~Meng and K.~T.~Chao,
Coupled-Channel and Screening Effects in Charmonium Spectrum,
Phys. Rev. D \textbf{80}, 014012 (2009).

\bibitem{Ono:1983rd}
S.~Ono and N.~Tornqvist,
Continuum Mixing and Coupled Channel Effects in $c \bar{c}$ and $b \bar{b}$ Quarkonium,
Z. Phys. C \textbf{23}, 59 (1984).

\bibitem{Hanhart}
  F.~K.~Guo, C.~Hanhart, Q.~Wang and Q.~Zhao,
  Could the near-threshold $XYZ$ states be simply kinematic effects?,
  Phys.\ Rev.\ D {\bf 91}, 051504 (2015).

\bibitem{Guorep}
  F.~K.~Guo, C.~Hanhart, U.~G.~Mei\ss ner, Q.~Wang, Q.~Zhao and B.~S.~Zou,
  Hadronic molecules,
  Rev.\ Mod.\ Phys.\  {\bf 90},  015004 (2018).

\bibitem{alberto}
  A.~Martinez Torres, K.~P.~Khemchandani, F.~S.~Navarra, M.~Nielsen and E.~Oset,
  The Role of $f_0(1710)$ in the $\phi \omega$ Threshold Peak of $J/\Psi \to \gamma \phi \omega$,
  Phys.\ Lett.\ B {\bf 719}, 388 (2013).

\bibitem{exp2}
  M.~Ablikim {\it et al.} [BES Collaboration],
  Observation of a near-threshold enhancement in the $\omega \phi$ mass spectrum from the doubly OZI suppressed decay $J / \psi \to \gamma \omega \phi$,
  Phys.\ Rev.\ Lett.\  {\bf 96}, 162002 (2006).




\bibitem{daniexp}
  D.~Gamermann and E.~Oset,
  Hidden charm dynamically generated resonances and the $e^+ e^- \to J / \psi D \bar{D}$, $J / \psi D \bar{D}^*$ reactions,
  Eur.\ Phys.\ J.\ A {\bf 36}, 189 (2008).






\bibitem{Uehara:2005qd} 
  S.~Uehara {\it et al.} [Belle Collaboration],
  Observation of a $\chi^\prime_{c2}$ candidate in $\gamma \gamma to D \bar D$ production at BELLE,
  Phys.\ Rev.\ Lett.\  {\bf 96}, 082003 (2006).

\bibitem{Aubert:2010ab} 
  B.~Aubert {\it et al.} [BaBar Collaboration],
  Observation of the $\chi_{c2}(2p)$ Meson in the Reaction $\gamma \gamma \to D \bar{D}$ at {BaBar},
  Phys.\ Rev.\ D {\bf 81}, 092003 (2010).

\bibitem{Aaij:2019evc}
R.~Aaij \textit{et al.} [LHCb],
Near-threshold $D\bar{D}$ spectroscopy and observation of a new charmonium state,
JHEP \textbf{07} (2019), 035.


\bibitem{Gamermann}
  D.~Gamermann, E.~Oset, D.~Strottman and M.~J.~Vicente Vacas,
  Dynamically generated open and hidden charm meson systems,
  Phys.\ Rev.\ D {\bf 76}, 074016 (2007).


\bibitem{Gamermann:2008jh}
D.~Gamermann, E.~Oset and B.~S.~Zou,
The Radiative decay of $\psi(3770)$ into the predicted scalar state $X(3700)$,
Eur. Phys. J. A \textbf{41} (2009), 85-91.

\bibitem{Bramon:1992kr}
A.~Bramon, A.~Grau and G.~Pancheri,
Intermediate vector meson contributions to $V_0\to  P^0 P^0 \gamma$ decays,
Phys. Lett. B \textbf{283} (1992), 416-420.



\bibitem{Dai}
  L.~R.~Dai, J.~J.~Xie and E.~Oset,
  $B^0 \rightarrow D^0 \bar{D}^0 K^0$ , $B^+ \rightarrow D^0 \bar{D}^0 K^+$ , and the scalar $D \bar{D}$ bound state,
  Eur.\ Phys.\ J.\ C {\bf 76},  121 (2016).


\bibitem{Watson:1952ji}
K.~M.~Watson,
The Effect of final state interactions on reaction cross-sections,
Phys. Rev. \textbf{88} (1952), 1163-1171.

\bibitem{migdal}
 A. B. Migdal, 
 The Theory of Nuclear Reactions with Production of Slow Particles,
Zh. \'Eksp. Teor. Fiz. {\bf 28}, no 1, 3 (1995). [English translation -Sov. Phys. JETP \textbf{1} (1955) 1, 2-6].

\bibitem{Hanhart:2003pg}
C.~Hanhart,
Meson production in nucleon-nucleon collisions close to the threshold,
Phys. Rept. \textbf{397} (2004), 155-256.

\bibitem{Hanhart:1998rn}
C.~Hanhart and K.~Nakayama,
On the treatment of $N N$ interaction effects in meson production in $N N$ collisions,
Phys. Lett. B \textbf{454} (1999), 176-180.



\bibitem{Sibirtsev:2004id}
A.~Sibirtsev, J.~Haidenbauer, S.~Krewald, U.~G.~Mei{\ss}ner and A.~W.~Thomas,
Near threshold enhancement of the $p\bar{p}$ mass spectrum in $J/\psi$ decay,
Phys. Rev. D \textbf{71}, 054010 (2005).

\bibitem{Baru:2000hg}
V.~Baru, A.~M.~Gasparian, J.~Haidenbauer, A.~E.~Kudryavtsev and J.~Speth,
On the Migdal-Watson approach to FSI effects in meson production in $N N$ collisions,
Phys. Atom. Nucl. \textbf{64}, 579-584 (2001).

\bibitem{Haidenbauer:2005eh}
J.~Haidenbauer, S.~Krewald, U.~G.~Mei{\ss}ner, A.~Sibirtsev and A.~W.~Thomas,
Analysis of the $p\bar{p}$  mass spectrum from $J/\psi$ decay,
AIP Conf. Proc. \textbf{796} (2005) no.1, 137-140.

\bibitem{Oset:2016lyh}
E.~Oset, W.~H.~Liang, M.~Bayar, J.~J.~Xie, L.~R.~Dai, M.~Albaladejo, M.~Nielsen, T.~Sekihara, F.~Navarra, L.~Roca, M.~Mai, J.~Nieves, J.~M.~Dias, A.~Feijoo, V.~K.~Magas, A.~Ramos, K.~Miyahara, T.~Hyodo, D.~Jido, M.~Döring, R.~Molina, H.~X.~Chen, E.~Wang, L.~Geng, N.~Ikeno, P.~Fernández-Soler and Z.~F.~Sun,
Weak decays of heavy hadrons into dynamically generated resonances,
Int. J. Mod. Phys. E \textbf{25}, 1630001 (2016).






\bibitem{Hanhart:2015zyp}
  C.~Hanhart,
  Amplitude Analysis for Mesons and Baryons: Tools and Technology, Talk at the Hadron 2015 Conference,
  AIP Conf.\ Proc.\  {\bf 1735},  020015 (2016)

\bibitem{Hyodo:2020czb}
T.~Hyodo and M.~Niiyama,
QCD and the Strange Baryon Spectrum,
[arXiv:2010.07592 [hep-ph]].


\bibitem{Wang:2020pyy}
G.~Y.~Wang, L.~Roca, E.~Wang, W.~H.~Liang and E.~Oset,
Signatures of the two $K_1(1270)$ poles in $D^+\to \nu e^+ V P$ decay,
Eur. Phys. J. C \textbf{80} (2020) no.5, 388.

\bibitem{Wang:2019mph}
G.~Y.~Wang, L.~Roca and E.~Oset,
Discerning the two $K_1(1270)$ poles in $D^0\to \pi^+ V P$ decay,
Phys. Rev. D \textbf{100} (2019) no.7, 074018.


\bibitem{Dong:2020hxe}
X.~K.~Dong, F.~K.~Guo and B.~S.~Zou,
Why there are many threshold structures in hadron spectrum with heavy quarks,
[arXiv:2011.14517 [hep-ph]].

\bibitem{Wang:2020elp}
E.~Wang, H.~S.~Li, W.~H.~Liang and E.~Oset,
Analysis of the $ {\gamma\gamma \to D\bar D}$ reaction and the $D\bar{D}$ bound state,
[arXiv:2010.15431 [hep-ph]].


\bibitem{isgur}
  J.~D.~Weinstein and N.~Isgur,
  $K \bar K$ Molecules,
  Phys.\ Rev.\ D {\bf 41}, 2236 (1990).

\bibitem{npa}
  J.~A.~Oller and E.~Oset,
  Chiral symmetry amplitudes in the S wave isoscalar and isovector channels and the $\sigma$, f$_0$(980), a$_0$(980) scalar mesons,
  Nucl.\ Phys.\ A {\bf 620}, 438 (1997)
  Erratum: [Nucl.\ Phys.\ A {\bf 652}, 407 (1999)].

\bibitem{Kaiser}
  N.~Kaiser,
  $\pi \pi$ $S$ wave phase shifts and nonperturbative chiral approach,
  Eur.\ Phys.\ J.\ A {\bf 3}, 307 (1998).

\bibitem{Markushin}
  M.~P.~Locher, V.~E.~Markushin and H.~Q.~Zheng,
  Structure of $f_0 (980)$ from a coupled channel analysis of $S$ wave $\pi \pi$ scattering,
  Eur.\ Phys.\ J.\ C {\bf 4}, 317 (1998).

\bibitem{Nieves}
  J.~Nieves and E.~Ruiz Arriola,
  Bethe-Salpeter approach for unitarized chiral perturbation theory,
  Nucl.\ Phys.\ A {\bf 679}, 57 (2000).

\bibitem{Nieves:2012tt}
  J.~Nieves and M.~P.~Valderrama,
  The Heavy Quark Spin Symmetry Partners of the $X(3872)$,
  Phys.\ Rev.\ D {\bf 86}, 056004 (2012).

\bibitem{HidalgoDuque:2012pq}
  C.~Hidalgo-Duque, J.~Nieves and M.~P.~Valderrama,
  Light flavor and heavy quark spin symmetry in heavy meson molecules,
  Phys.\ Rev.\ D {\bf 87}, 076006 (2013).

\bibitem{Prelovsek:2020eiw}
S.~Prelovsek, S.~Collins, D.~Mohler, M.~Padmanath and S.~Piemonte,
Charmonium-like resonances with $J^{PC}=0^{++},2^{++}$ in coupled $D\bar D$, $D_s\bar D_s$ scattering on the lattice,
[arXiv:2011.02542 [hep-lat]].

\bibitem{ddbarh}
Talk given by Johnson, Daniel (CERN), $B\to D\bar{D}h$ decays: a new (virtual) laboratory for exotic particle searches at LHCb, LHCb-TALK-2020-148, CERN-LHC Seminar, CERN, Switzerland, 11-Aug-2020, see http://cds.cern.ch/record/2727764

\bibitem{Tim}
Tim Gershon, private communication.

\bibitem{Aaij:2020ypa}
R.~Aaij \textit{et al.} [LHCb],
Amplitude analysis of the $B^+\to D^+D^-K^+$ decay,
Phys. Rev. D \textbf{102} (2020), 112003.


\bibitem{Aaij:2020hon}
R.~Aaij \textit{et al.} [LHCb],
A model-independent study of resonant structure in $B^+\to D^+D^-K^+$ decays,
Phys. Rev. Lett. \textbf{125} (2020), 242001.


\end{thebibliography}
  \end{document}